\documentclass[aps,PRB,twocolumn,superscriptaddress,preprintnumbers]{revtex4-2}
\usepackage[T1]{fontenc}
\usepackage{times}
\usepackage{graphicx,color}
\usepackage{amsfonts,amsmath,amssymb,amsbsy}
\usepackage[colorlinks=true, linkcolor=blue, citecolor=blue, urlcolor=blue]{hyperref}
\usepackage{float}

\begin{document}

\title{Third-order and fifth-order nonlinear spin-current generation in $g$%
-wave and $i$-wave altermagnets, and perfectly nonreciprocal spin-current in 
$f$-wave magnets}
\author{Motohiko Ezawa}
\affiliation{Department of Applied Physics, The University of Tokyo, 7-3-1 Hongo, Tokyo
113-8656, Japan}

\begin{abstract}
A prominent feature of $d$-wave altermagnets is the pure spin current
generated in the absence of spin-orbit interactions. In the context of
symmetry, there are the $s$-wave, the $p$-wave, the $d$-wave, the $f$-wave,
the $g$-wave and the $i$-wave magnets. In this paper, making an analytic
study of two-band Hamiltonian systems coupled with electrons, we demonstrate
unexpectedly that only the $\ell $-th order nonlinear transverse spin
current proportional to $E^{\ell }$ is generated in higher-wave symmetric
magnets when the number of the nodes is $\ell +1$. Here $E$ is applied
electric field. The nonlinear spin current is essential provided the linear
spin current is absent. Indeed, only the third-order nonlinear spin current
is generated in $g$-wave altermagnets, while only the fifth-order spin
current is generated in $i$-wave altermagnets. In particular, only the
second-order nonlinear spin current is generated in $f$-wave magnets, which
leads to a perfect nonreciprocal spin current. On the other hand, there is
no spin-current generation in $p$-wave magnets.
\end{abstract}

\date{\today }
\maketitle

\section{Introduction}

Pure spin current is a fundamental concept in spintronics, where spin
current flows but no charge current flows. It is usually generated by the
spin-Hall effect induced by spin-orbit interactions\cite{Dya,Dya2,Sinova}.
However, the spin-orbit interaction shortens the coherent length of the spin
current because it rotates the spin direction. It limits the application of
spin current. Recently, it was shown\cite{Naka,Gonza,NakaB,Bose,NakaRev}
that the spin current is generated without the spin-orbit interaction in $d$%
-wave altermagnets with the aid of a spin-split band structure. Its
coherence length may be longer than that generated by the spin-orbit
interaction.

In the context of symmetry, there are the $s$-wave, the $p$-wave\cite%
{Hayami,pwave,He}, the $d$-wave\cite{Naka,Gonza,NakaB,Hayami,Bose,NakaRev},
the $f$-wave\cite{pwave,He}, the $g$-wave\cite{SmejX,SmejX2} and the $i$-wave%
\cite{SmejX,SmejX2} magnets which are compatible with the lattice symmetry.
We call them higher-wave symmetric magnets. Especially, the $d$-wave, the $g$%
-wave and the $i$-wave magnets break time-reversal symmetry and called
altermagnets\cite{SmejX}. However, the problem of spin-current generation
still remains to be uncovered in the higher-wave symmetric magnets except
for the $d$-wave altermagnet.

There are several studies on nonlinear conductivity. Especially, the
second-order nonlinear conductivity has been extensively investigated\cite%
{Gao,Sodeman,Ideue,HLiu,Michishita,Watanabe,CWang,Oiwa,AGao,NWang,KamalDas,Kaplan,Ohmic,Xiang,ZGong}%
. The $\ell $-th nonlinear conductivity $\sigma ^{a_{1}a_{2}\cdots a_{\ell
};b}$ is defined by $j_{b}^{\ell }=\sigma ^{a_{1}a_{2}\cdots a_{\ell
};b}E_{a_{1}}E_{a_{2}}\cdots E_{a_{\ell }}$, where $E_{a_{i}}$ is the
applied electric field along the $a_{i}$ direction, $j_{b}^{\ell }$ is the $%
\ell $-th nonlinear current along the $b$ direction. Here, some directions
can be identical such as $a_{1}=a_{2}=\cdots =a_{\ell }=x$. Nonlinear spin
current generation is also interesting\cite%
{Hamamoto,Kameda,Hayami22B,Hayami24B}.

In this paper, we study linear and nonlinear spin-current generation in
higher-wave symmetric magnets and show that the system has only the $\ell $%
-th order nonlinear spin-Drude conductivity when the number of the node is $%
\ell +1$. Note that the linear spin current corresponds to the case of $\ell
=1$. We obtain the following results: There is no spin current generation in
the $s$-wave and $p$-wave magnets. Only the linear spin current is generated
in the $d$-wave altermagnet. The second-order nonlinear spin current is
generated in the $f$-wave magnet, where the current flowing direction is
identical irrespective of the direction of applied electric field. The
third-order nonlinear spin current is generated in the $g$-wave altermagnet,
while the fifth-order nonlinear spin current is generated in the $i$-wave
altermagnet. These results are summarized in the following table.

\begin{tabular}{|c|c|c|c|c|c|c|}
\hline
& $s$ & $p$ & $d$ & $f$ & $g$ & $i$ \\ \hline
nodes & 0 & 1 & 2 & 3 & 4 & 6 \\ \hline
$\ell $ &  &  & linear & 2nd NL & 3rd NL & 5th NL \\ \hline
2D & None & None & $\sigma _{\text{spin}}^{y;x}$ & $\sigma _{\text{spin}%
}^{xx;y}$ & $\sigma _{\text{spin}}^{yyy;x}$ & $\sigma _{\text{spin}%
}^{yyyyy;x}$ \\ \hline
3D & None & None & $\sigma _{\text{spin}}^{y;x}$ & $\sigma _{\text{spin}%
}^{zy;x}$ & $\sigma _{\text{spin}}^{xxx;z}$ & $\sigma _{\text{spin}%
}^{yyyyx;x}$ \\ \hline
\end{tabular}

\begin{figure*}[t]
\centerline{\includegraphics[width=0.98\textwidth]{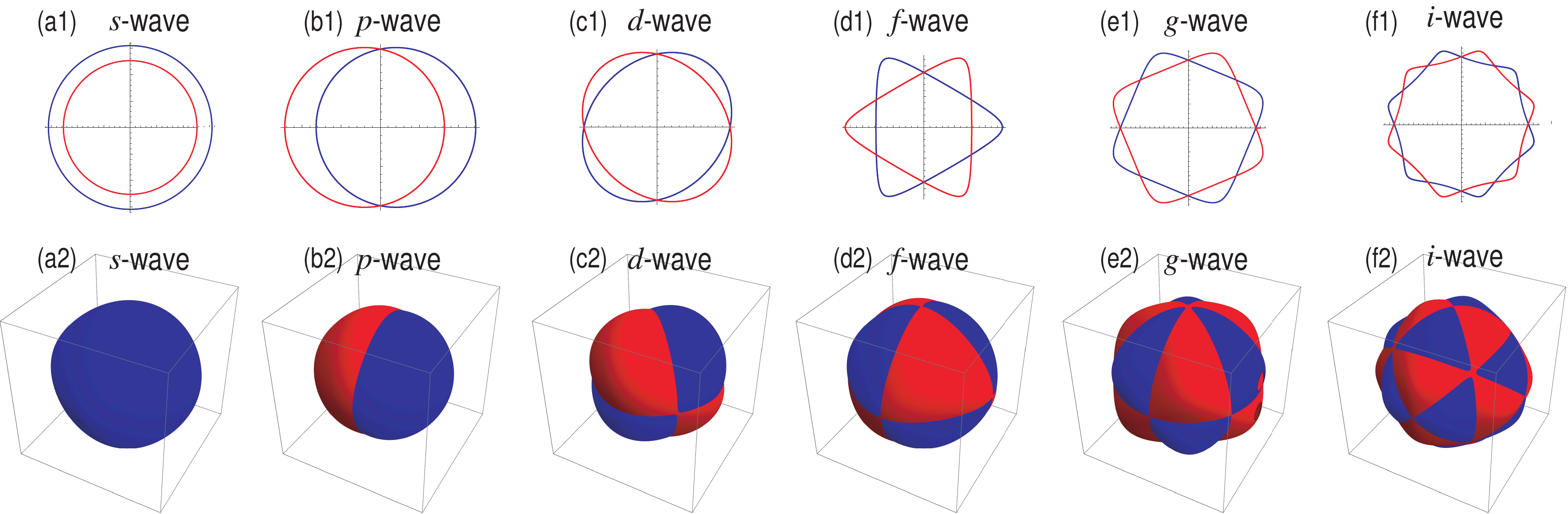}}
\caption{Fermi surfaces in two and three dimensions. (a1), (a2) $s$-wave
magnet; (b1), (b2) $p$-wave magnet; (c1), (c2) $d$-wave magnet; (d1), (d2) $%
f $-wave magnet; (e1), (e2) $g$-wave magnet; (f1) and (f2) $i$-wave magnet.
Red (blue) curves indicate up (down)-spin Fermi surfaces.}
\label{FigSurface}
\end{figure*}

\section{Fermi surface symmetry}

The Fermi surface of electrons coupled with a magnet is known to have the $s$%
-wave symmetry as shown in Fig.\ref{FigSurface}(a). Recently proposed
altermagnets generalize it to Fermi surfaces possessing higher-wave
symmetries\cite{SmejX,SmejX2}. The Fermi surface has $0,1,2,3,4$ and $6$\
nodes for the $s$-wave, $p$-wave, $d$-wave, $f$-wave, $g$-wave and $i$-wave
symmetry, respectively. The $d$-wave altermagnet has the Fermi surface with
the $d$-wave symmetry as shown in Fig.\ref{FigSurface}(c1). In the similar
way, the Fermi surface of the $g$-wave altermagnet is shown in Fig.\ref%
{FigSurface}(e1) and that of the $i$-wave alatermagnet is shown in Fig.\ref%
{FigSurface}(f1). Altermagnets break time-reversal symmetry. On the other
hand, $p$-wave magnets preserve time-reversal symmetry. Its Fermi surface
has the $p$-wave symmetry as shown in Fig.\ref{FigSurface}(b1). In a similar
way, $f$-wave magnets have Fermi surfaces with the $f$-wave symmetry as
shown in Fig.\ref{FigSurface}(d1). We note that there are no $h$-wave
magnets because of the incompatibility between the five-fold rotational
symmetry and the lattice symmetry. The simplest expressions on magnetic
terms with higher symmetries in two dimensions are summarized as follows.%
\begin{align}
H_{s}^{\text{2D}}& =J\sigma _{z},  \label{EqX} \\
H_{p}^{\text{2D}}& =Jk_{x}\sigma _{z}, \\
H_{d}^{\text{2D}}& =Jk_{x}k_{y}\sigma _{z}, \\
H_{f}^{\text{2D}}& =Jk_{x}\left( k_{x}^{2}-3k_{y}^{2}\right) \sigma _{z}, \\
H_{g}^{\text{2D}}& =Jk_{x}k_{y}\left( k_{x}^{2}-k_{y}^{2}\right) \sigma _{z},
\\
H_{i}^{\text{2D}}& =Jk_{x}k_{y}\left( 3k_{x}^{2}-k_{y}^{2}\right) \left(
k_{x}^{2}-3k_{y}^{2}\right) \sigma _{z}.
\end{align}%
Their tight-binding representation is constructed by replacing $k_{j}\mapsto
\sin ak_{j}$, $k_{j}^{2}\mapsto 2\left( 1-\cos ak_{j}\right) $\ with $j=x,y$
and the lattice constant $a$.

So far, we have discussed two-dimensional magnets. In the similar way, there
are three-dimensional magnets with higher-wave symmetries\cite{SmejX,SmejX2}
as summarized in Fig.\ref{FigSurface}(a2)\symbol{126}(f2). The simplest
expressions on magnetic terms with higher symmetries in three dimensions are
summarized as follows. 
\begin{align}
H_{d}^{\text{3D}}& =Jk_{z}\left( k_{x}+k_{y}\right) \sigma _{z}, \\
H_{f}^{\text{3D}}& =Jk_{x}k_{y}k_{z}\sigma _{z}, \\
H_{g}^{\text{3D}}& =Jk_{z}k_{x}\left( k_{x}^{2}-3k_{y}^{2}\right) \sigma
_{z}, \\
H_{i}^{\text{3D}}& =J\left( k_{x}^{2}-k_{y}^{2}\right) \left(
k_{y}^{2}-k_{z}^{2}\right) \left( k_{z}^{2}-k_{x}^{2}\right) \sigma _{z}.
\label{EqY}
\end{align}%
Their tight-binding representation is constructed by replacing $k_{j}\mapsto
\sin ak_{j}$, $k_{j}^{2}\mapsto 2\left( 1-\cos ak_{j}\right) $\ with $j=x,y,z
$.

\section{Nonlinear Drude conductivity}

The Drude conductivity is the linear conductivity. It was generalized to the
second-order nonlinear Drude conductivity\cite{Ideue,NLDrude}. The
third-order nonlinear Drude conductivity was also derived\cite{YFang}. They
are derived in the static limit. We derive the $\ell $-th order nonlinear
Drude conductivity for a single band system for an arbitrary $\ell $, which
is valid even in the nonstatic regime. In the clean limit, the Drude
conductivity is dominant.

The current is given by%
\begin{equation}
\mathbf{j}=-e\int d^{D}kf\mathbf{v},  \label{currentJ}
\end{equation}%
where $D$\ is the dimension; $f$ is the Fermi distribution function in the
presence of $\mathbf{E}$; $\mathbf{v}$ is the velocity,%
\begin{equation}
\mathbf{v}=\frac{1}{\hbar }\frac{\partial \varepsilon }{\partial \mathbf{k}},
\label{Velocity}
\end{equation}%
where $\varepsilon $ is the energy of the Hamiltonian and we have dropped
the anomalous velocity term proportional to $-e\mathbf{E}\times \mathbf{%
\Omega }/\hbar $ because the Berry curvature $\mathbf{\Omega }$ is zero due
to the single band condition.

Let us present formulas in two dimensions. The generalization to three
dimensions is straightforward. We expand the current in terms of electric
field $E$ as%
\begin{equation}
j_{b}=\sum_{\ell _{1},\ell _{2}=0}\sigma ^{x^{\ell _{1}}y^{\ell
_{2}};b}(E_{x})^{\ell _{1}}(E_{y})^{\ell _{2}}.  \label{Expansion}
\end{equation}%
Then, the $(\ell _{1}+\ell _{2})$-th order conductivity is defined by

\begin{equation}
\sigma ^{x^{\ell _{1}}y^{\ell _{2}};b}=\frac{1}{\ell _{1}!\ell _{2}!}\frac{%
\partial ^{\ell _{1}+\ell _{2}}j_{b}}{\partial E_{x}^{\ell _{1}}\partial
E_{y}^{\ell _{2}}}.
\end{equation}%
The semi-classical Boltzmann equation in the presence of electric field $%
\mathbf{E}$ is given by%
\begin{equation}
\partial _{t}f-\frac{e\mathbf{E}}{\hbar }\cdot \nabla _{\mathbf{k}}f=-\frac{%
f-f^{\left( 0\right) }}{\tau },
\end{equation}%
where $\tau $\ is the relaxation time and $f^{\left( 0\right) }$ is the
Fermi distribution function at the equilibrium with the chemical potential $%
\mu $,%
\begin{equation}
f^{\left( 0\right) }=1/\left( \exp \left( \varepsilon -\mu \right) +1\right)
.
\end{equation}%
Corresponding to Eq.(\textsl{\ref{Expansion}}), we expand the Fermi
distribution in powers of $\mathbf{E}$,%
\begin{equation}
f=f^{\left( 0\right) }+f^{\left( 1\right) }+\cdots .  \label{f}
\end{equation}%
The recursive solution gives\cite{YFang}%
\begin{equation}
f^{\left( \ell _{1}+\ell _{2}\right) }=\left( \frac{e/\hbar }{i\omega
+1/\tau }\right) ^{\ell _{1}+\ell _{2}}\frac{\partial ^{\ell _{1}+\ell
_{2}}f^{\left( 0\right) }}{\partial k_{x}^{\ell _{1}}\partial k_{y}^{\ell
_{2}}}(E_{x})^{\ell _{1}}(E_{x})^{\ell _{2}},
\end{equation}%
where $\omega $ is the frequency of the applied electric field $\mathbf{E}$.
The current (\ref{currentJ}) is expanded as in Eq.(\ref{f}) with 
\begin{align}
j_{b}^{\left( \ell _{1}+\ell _{2}\right) }& =-e\int d^{D}kf^{\left( \ell
_{1}+\ell _{2}\right) }v_{b}  \notag \\
& =-\frac{e}{\hbar }\left( \frac{e/\hbar }{i\omega +1/\tau }\right) ^{\ell
_{1}+\ell _{2}}  \notag \\
& \qquad \quad \times \int d^{D}k\,\frac{\partial \varepsilon }{\partial
k_{b}}\frac{\partial ^{\ell _{1}+\ell _{2}}f^{\left( 0\right) }}{\partial
k_{x}^{\ell _{1}}\partial k_{y}^{\ell _{2}}}(E_{x})^{\ell _{1}}(E_{x})^{\ell
_{2}}.
\end{align}%
The $(\ell _{1}+\ell _{2})$-th nonlinear Drude conductivity is defined by%
\begin{equation}
\sigma ^{x^{\ell _{1}}y^{\ell _{2}};b}=\frac{\left( -e/\hbar \right) ^{\ell
_{1}+\ell _{2}+1}}{\left( i\omega +1/\tau \right) ^{\ell _{1}+\ell _{2}}}%
\int d^{D}kf^{\left( 0\right) }\frac{\partial ^{\ell _{1}+\ell
_{2}+1}\varepsilon }{\partial k_{x}^{\ell _{1}}\partial k_{y}^{\ell
_{2}}\partial k_{b}}.
\end{equation}%
The static limit is obtained simply by setting $\omega =0$ in this equation.

In this paper, we consider a Hamiltonian, where the spin is a good quantum
number, $\sigma _{z}=s=\pm 1$. We can define the $\ell $-th order
spin-dependent Drude conductivity for each spin $s$ by the formula%
\begin{equation}
\sigma _{s}^{x^{\ell _{1}}y^{\ell _{2}};b}=\frac{\left( -e/\hbar \right)
^{\ell _{1}+\ell _{2}+1}}{\left( i\omega +1/\tau \right) ^{\ell _{1}+\ell
_{2}}}\int d^{D}k\,f_{s}^{\left( 0\right) }\frac{\partial ^{\ell _{1}+\ell
_{2}+1}\varepsilon _{s}}{\partial k_{x}^{\ell _{1}}\partial k_{y}^{\ell
_{2}}\partial k_{b}},  \label{Drude}
\end{equation}%
where $s=\uparrow \downarrow $. Let us use $s=\uparrow \downarrow $\ within
indices and $s=\pm 1$\ in equations. This formula is nontrivial only when%
\begin{equation}
\frac{\partial ^{\ell _{1}+\ell _{2}+1}\varepsilon _{s}}{\partial
k_{x}^{\ell _{1}}\partial k_{y}^{\ell _{2}}\partial k_{b}}\neq 0,
\label{BasicCond}
\end{equation}%
which leads \ to a conclusion that there is no $\ell $-th order nonlinear
spin-Drude conductivity for $\ell \geq \ell _{1}+\ell _{2}$. It is necessary
to calculate explicitly the $\ell $-th order nonlinear spin-Drude
conductivity for $\ell =0,1,\cdots ,\ell _{1}+\ell _{2}-1$. In particular,
the choice of $\ell =0$\ and $1$\ yield the persistent spin current without
electric field and the linear spin conductivity, respectively.

We define the $(\ell _{1}+\ell _{2})$-th order nonlinear spin-Drude
conductivity by 
\begin{equation}
\sigma _{\text{spin}}^{x^{\ell _{1}}y^{\ell _{2}};b}=\frac{\sigma _{\uparrow
}^{x^{\ell _{1}}y^{\ell _{2}};b}-\sigma _{\downarrow }^{x^{\ell _{1}}y^{\ell
_{2}};b}}{2}.  \label{spinDr}
\end{equation}%
On the other hand, the charge conductivity is given by%
\begin{equation}
\sigma _{\text{charge}}^{x^{\ell _{1}}y^{\ell _{2}};b}=\sigma _{\uparrow
}^{x^{\ell _{1}}y^{\ell _{2}};b}+\sigma _{\downarrow }^{x^{\ell _{1}}y^{\ell
_{2}};b}.  \label{ChargeCurre}
\end{equation}%
In the following sections, we calculate Eq.(\ref{Drude}) for various
higher-wave symmetric magnets.

In general, there are other contributions to the nonlinear conductivity from
the quantum metric and the Berry curvature dipole for multi-band system\cite%
{Gao,Sodeman,Ideue,HLiu,Michishita,Watanabe,CWang,Oiwa,AGao,NWang,KamalDas,Kaplan,Ohmic,Xiang,ZGong}%
. In the present model, the Hamiltonian is diagonal with respect to the spin
degrees of freedom, and hence, the system is an essentially single-band
system. Hence, there is no contribution in the nonlinear conductivity from
the quantum metric and the Berry curvature dipole.

The kinetic energy of free fermions is described by the Hamiltonian%
\begin{equation}
H_{0}^{2\text{D}}=\frac{\hbar ^{2}\left( k_{x}^{2}+k_{y}^{2}\right) }{2m}%
\sigma _{0}  \label{Kine2D}
\end{equation}%
in two dimensions, and%
\begin{equation}
H_{0}^{3\text{D}}=\frac{\hbar ^{2}\left(
k_{x}^{2}+k_{y}^{2}+k_{z}^{2}\right) }{2m}\sigma _{0}  \label{Kine3D}
\end{equation}%
in three dimensions, where $m$ is the free-fermion mass, and $\sigma _{0}$
is the $2\times 2$ identity matrix.

The linear longitudinal charge conductivity is always present due to the
kinetic Hamiltonian (\ref{Kine2D}) or (\ref{Kine3D}),%
\begin{equation}
\sigma _{\text{charge}}^{x;x}=\frac{\left( e/\hbar \right) ^{2}}{i\omega
+1/\tau }\left( V_{\uparrow }^{\text{F}}+V_{\downarrow }^{\text{F}}\right) ,
\end{equation}%
where $V_{s}^{\text{F}}$ is the spin-dependent Fermi volume,%
\begin{equation}
V_{s}^{\text{F}}\equiv \int d^{D}kf_{s}^{\left( 0\right) },
\end{equation}%
and we have used the relation (\ref{Drude}), or%
\begin{equation}
\sigma _{s}^{x;x}=\frac{\left( e/\hbar \right) ^{2}}{i\omega +1/\tau }\int
d^{D}k\frac{\partial ^{2}\varepsilon }{\partial k_{x}^{2}}f_{s}^{\left(
0\right) }=\frac{\left( e/\hbar \right) ^{2}}{i\omega +1/\tau }V_{s}^{\text{F%
}}.
\end{equation}

On the other hand, the magnetic terms with higher symmetries generate
transverse spin currents. We investigate linear and nonlinear transverse
spin conductivities by taking the input and output along the $x$, $y$\ and $%
z $\ axis.

\section{$s$-wave magnet in 2D and 3D}

We consider the Hamiltonian of the $s$-wave magnet given by

\begin{equation}
H^{\text{total}}=H_{0}+H_{s},
\end{equation}%
where $H_{0}=H_{0}^{2\text{D}}$ in two dimensions and $H_{0}=H_{0}^{3\text{D}%
}$ in three dimensions with%
\begin{equation}
H_{s}=J\sigma _{z}.  \label{EqA}
\end{equation}%
This system describes a ferromagnet.

The Fermi surface is determined by solving%
\begin{equation}
\frac{\hbar ^{2}k^{2}}{2m}+sJ=\mu ,
\end{equation}%
which is a circle with the radius $k_{s}^{\text{F}}$ depending on the spin $%
s $,%
\begin{equation}
\hbar k_{s}^{\text{F}}=\sqrt{2m\left( \mu -sJ\right) }.
\end{equation}%
The Fermi volume is 
\begin{equation}
V_{s}^{\text{F}}=\frac{\pi }{\hbar ^{2}}(2m\mu -sJ)  \label{FS-s}
\end{equation}%
in two dimensions and%
\begin{equation}
V_{s}^{\text{F}}=\frac{4\pi }{3\hbar ^{3}}(2m\mu -sJ)^{3/2}
\end{equation}%
in three dimensions. They are shown in Fig.\ref{FigSurface}(a1) and (a2).

It follows from the condition (\ref{BasicCond}) that there is no $\ell $-th
order spin-Drude conductivity for all $\ell \geq 2$.\textsl{\ }We examine
explicitly the spin conductivity for $\ell =0,1$ in the two-dimensional
system. We examine all the nontrivial ones in the condition (\ref{BasicCond}%
),%
\begin{align}
\frac{\partial \varepsilon _{s}}{\partial k_{x}} =&\frac{\hbar ^{2}k_{x}}{m}%
,\quad \frac{\partial \varepsilon _{s}}{\partial k_{y}}=\frac{\hbar ^{2}k_{y}%
}{m},  \notag \\
\frac{\partial ^{2}\varepsilon _{s}}{\partial k_{x}^{2}} =&\frac{\partial
^{2}\varepsilon _{s}}{\partial k_{y}^{2}}=\frac{\hbar ^{2}}{m},
\end{align}%
which arise from the kinetic term (\ref{Kine2D}).

We examine the spin current for $\ell =0$, which corresponds to the
persistent spin current without electric field. It is given by%
\begin{equation}
j_{x}^{\left( 0\right) }=-\frac{e}{\hbar }\int d^{2}k\frac{\partial
\varepsilon }{\partial k_{x}}f_{s}^{\left( 0\right) }=-\frac{e\hbar }{m}\int
d^{2}k\,k_{x}f_{s}^{\left( 0\right) }=0,
\end{equation}%
because the integrand is an odd function of $k_{x}$.

We next examine the linear spin conductivity with $\ell =1$. It is given by%
\begin{equation}
\sigma _{s}^{x;x}=\frac{e}{\hbar }\int d^{2}k\frac{\partial ^{2}\varepsilon 
}{\partial k_{x}^{2}}f_{s}^{\left( 0\right) }=\frac{e\hbar }{m}V_{s}^{\text{F%
}},
\end{equation}%
where $V_{s}^{\text{F}}$ is the spin-dependent Fermi volume (\ref{FS-s}).
Hence, we obtain,%
\begin{equation}
\sigma _{\text{spin}}^{x;x}=\frac{\sigma _{\uparrow }^{x;x}-\sigma
_{\downarrow }^{x;x}}{2}=\frac{e\hbar }{m}\frac{V_{\uparrow }^{\text{F}%
}-V_{\downarrow }^{\text{F}}}{2},
\end{equation}%
implying that there follows a spin polarized current in ferromagnet, where
both charge and spin currents flow. The three-dimensional system is
similarly analyzed with similar conclusion.

\section{$p$-wave magnet in 2D}

We consider the Hamiltonian of the $p$-wave magnet in two dimensions given by%
\begin{equation}
H^{\text{total}}=H_{0}^{2\text{D}}+H_{p}.  \label{pModel}
\end{equation}%
The second term represents the $p$-wave magnet\cite%
{pwave,Okumura,Maeda,EzawaPwave,Brek,EzawaPNeel},%
\begin{equation}
H_{p}=J\sigma _{z}k_{x}.
\end{equation}%
The spin-dependent Hamiltonian is explicitly written as

\begin{align}
H_{s}^{\text{total}} =&\frac{\hbar ^{2}\left( k_{x}^{2}+k_{y}^{2}\right) }{2m%
}+sJk_{x}  \notag \\
=&\frac{\hbar ^{2}}{2m}\left( k_{x}+\frac{msJ}{\hbar ^{2}}\right) ^{2}+\frac{%
\hbar ^{2}k_{y}^{2}}{2m}-\frac{mJ^{2}}{2\hbar ^{2}}.
\end{align}%
The Fermi surface is a circle for each spin $s$ as shown in Fig.\ref%
{FigSurface}(b1).

We make a change of variable from $k_{x}$ to $k_{x}^{\prime }$,%
\begin{equation}
k_{x}^{\prime }\equiv k_{x}+\frac{msJ}{\hbar ^{2}}.  \label{ChangeK}
\end{equation}%
The Hamiltonian is rewritten as%
\begin{equation}
H_{s}^{\text{total}}=\frac{\hbar ^{2}\left( k_{x}^{\prime
2}+k_{y}^{2}\right) }{2m}-\frac{mJ^{2}}{2\hbar ^{2}},
\end{equation}%
which has the form of the free fermion. The Fermi surface is described by%
\begin{equation}
\hbar k^{\prime \text{F}}=\sqrt{2m\left( \mu +\frac{mJ^{2}}{2\hbar ^{2}}%
\right) },  \label{pFS}
\end{equation}%
and the Fermi volume is%
\begin{equation}
V^{\text{F}}=\frac{2\pi m}{\hbar ^{2}}\left( \mu +\frac{mJ^{2}}{2\hbar ^{2}}%
\right) .
\end{equation}%
Because both the Hamiltonian and the Fermi surfaces are independent of the
spin $s$ in the $(k_{x}^{\prime },k_{y})$ space,\ there is no $\ell $-th
order spin-Drude conductivity (\ref{spinDr}) for all $\ell \geq 0$.

A comment is in order. The Fermi surfaces of the $p$-wave magnet are
peculiar in comparison with all others in Fig.\ref{FigSurface}, where the
Fermi surfaces are shifted oppositely for the opposite spins, as shown in
Fig.\ref{FigSurface}(b2). Hence, one might assume that there must be a net
spin current with $\ell =0$, $\sigma _{\text{spin}}^{;x}\neq 0$, implying
the existence of a persistent spin current without electric field. This
intuitive picture would be correct if fermions were free fermions, where 
\begin{equation}
v_{x}=\hbar k_{x}/m.  \label{EqE}
\end{equation}%
Then, because $f_{s}^{\left( 0\right) }k_{x}^{\prime }$ is an odd function
of $k_{x}^{\prime }$, the current would be%
\begin{align}
j_{x}^{\left( 0\right) } &=-e\int d^{2}kf_{s}^{\left( 0\right) }v_{x}=-\frac{%
e\hbar }{m}\int d^{2}k^{\prime }f_{s}^{\left( 0\right) }(k_{x}^{\prime }-%
\frac{sJ}{\hbar ^{2}})  \notag \\
&=sJ\frac{e}{m\hbar }V^{\text{F}},
\end{align}%
where $V^{\text{F}}$ is the Fermi volume. It would imply the existence of
the persistent spin current without external electric field $E$.\ However,
this is not correct because fermions are not free fermions. The velocity is
determined by the formula (\ref{Velocity}), 
\begin{equation}
v_{x}=\hbar k_{x}/m+sJ/\hbar ,
\end{equation}%
and not by Eq.(\ref{EqE}). These two terms cancel each other to produce no
persistent spin current\ without external field $E$.

For the sake of completeness, let us calculate explicitly the persistent
spin-dependent current\ without external field $E$. The current is given by
Eq.(\ref{currentJ}) with the choice of $\ell =0$, which is calculated as%
\begin{align}
j_{x}^{\left( 0\right) }& =-e\int d^{2}kf_{s}^{\left( 0\right) }v_{x}=-\frac{%
e}{\hbar }\int d^{2}k^{\prime }\frac{\partial k_{x}^{\prime }}{\partial k_{x}%
}\frac{\partial \varepsilon _{s}}{\partial k_{x}^{\prime }}f_{s}^{\left(
0\right) }  \notag \\
& =-\frac{e}{\hbar }\int d^{2}k^{\prime }\frac{\partial \varepsilon _{s}}{%
\partial k_{x}^{\prime }}f_{s}^{\left( 0\right) }=-\frac{e\hbar }{m}\int
d^{2}k^{\prime }\,k_{x}^{\prime }f_{s}^{\left( 0\right) }=0.
\end{align}%
Hence, no zeroth spin current is generated,\ i.e., $\sigma _{\text{spin}%
}^{;x}=0$.\ 

Next, we calculate the linear spin-dependent conductivity ($\ell =1$). In
the $(k_{x}^{\prime },k_{y})$ space we obtain\ 
\begin{align}
\sigma _{s}^{x;x} &=\frac{e^{2}/\hbar ^{2}}{i\omega +1/\tau }\int
d^{2}k^{\prime }\frac{\partial ^{2}\varepsilon _{s}}{\partial k_{x}^{\prime
2}}f^{\left( 0\right) }  \notag \\
&=\frac{e^{2}/m}{i\omega +1/\tau }\int d^{2}k^{\prime }f^{\left( 0\right) }=%
\frac{e^{2}/m}{i\omega +1/\tau }V^{\text{F}}
\end{align}%
This is independent of the spin $s$, and hence, $\sigma _{\text{spin}%
}^{x;x}=0$. However, the linear charge conductivity is generated,%
\begin{equation}
\sigma _{\text{charge}}^{x;x}\equiv \sigma _{\uparrow }^{x;x}+\sigma
_{\downarrow }^{x;x}=\frac{4e\pi }{\hbar }\left( \mu +\frac{mJ^{2}}{2\hbar
^{2}}\right) .
\end{equation}%
It is interesting that the linear charge conductivity has a $J$ dependence.

The tight-binding model corresponding to the continuum model (\ref{pModel})
is given by 
\begin{equation}
H_{0}^{2\text{D}}=\frac{\hbar ^{2}}{ma^{2}}\left( 2-\cos ak_{x}-\cos
ak_{y}\right) \sigma _{0}+J\sigma _{z}\sin ak_{x},
\end{equation}%
which is defined on the square lattice.

\section{$p$-wave magnet in 3D}

We consider the Hamiltonian of a $p$-wave magnet in three dimensions given by%
\begin{equation}
H^{\text{total}}=H_{0}^{3\text{D}}+H_{p}.  \label{p3Model}
\end{equation}%
The second term represents the $p$-wave magnet,%
\begin{equation}
H_{p}=J\sigma _{z}k_{x}.
\end{equation}
The Fermi surface is shown in Fig.\ref{FigSurface}(b2).

Making a change of variable from $k_{x}$ to $k_{x}^{\prime }$ as in Eq.(\ref%
{ChangeK}), the Hamiltonian is rewritten as%
\begin{equation}
H_{s}=\frac{\hbar ^{2}\left( k_{x}^{\prime 2}+k_{y}^{2}+k_{z}^{2}\right) }{2m%
}-\frac{mJ^{2}}{2\hbar ^{2}},
\end{equation}%
which has the form of the free fermion. The Fermi surface is described by%
\begin{equation}
\hbar k^{\prime \text{F}}=\sqrt{2m\left( \mu +\frac{mJ^{2}}{2\hbar ^{2}}%
\right) }.
\end{equation}%
There is no $\ell $-th order spin-Drude conductivity for all $\ell \geq 0$
as in the two-dimensional system.\ The physics is essentially identical to
the $p$-wave magnet in two dimensions.

The tight-binding model corresponding to the continuum model (\ref{p3Model})
is given by 
\begin{align}
H_{0}^{2\text{D}}=& \frac{\hbar ^{2}}{ma^{2}}\left( 3-\cos ak_{x}-\cos
ak_{y}-\cos ak_{z}\right) \sigma _{0}  \notag \\
& +J\sin ak_{x}\sigma _{z},
\end{align}%
which is defined on the cubic lattice. 
\begin{figure}[t]
\centerline{\includegraphics[width=0.48\textwidth]{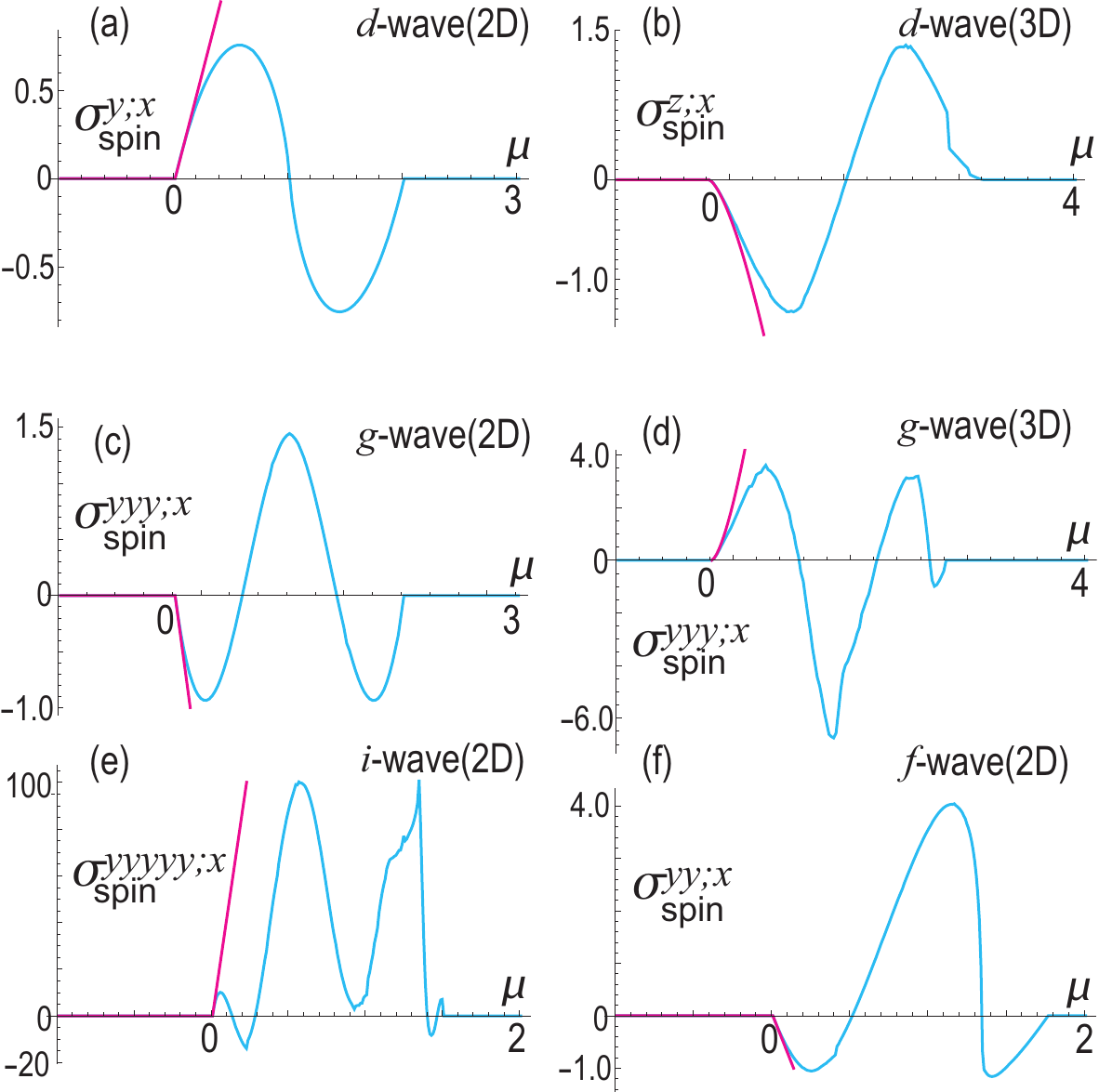}}
\caption{$\protect\mu $ dependence of the spin-Drude conductivity in units
of $e^{\ell }\protect\tau ^{\ell -1}\protect\varepsilon _{0}/\left( \hbar
^{\ell }k_{0}\right) $. (a) The $d$-wave magnet ($\ell =1$) in 2D, (b) the $%
d $-wave magnet ($\ell =1$) in 3D, (c) the $g$-wave magnet ($\ell =3$) in
2D, (d) the $g$-wave magnet ($\ell =3$) in 3D, (e) the $i$-wave magnet ($%
\ell =5$) in 2D, (f) the $f$-wave magnet ($\ell =2$) in 2D. Cyan curves
represent numerical results based on tight-binding models, while magenta
curves represent analytical results based on continuum models. The horizontal axis is $\mu$ in units of $\varepsilon _{0}$.We have set $%
\hbar ^{2}/\left( ma^{2}\right) =\protect\varepsilon _{0}/4$ and $J=0.1%
\protect\varepsilon _{0}$.}
\label{FigEyyyx}
\end{figure}

\begin{figure}[t]
\centerline{\includegraphics[width=0.48\textwidth]{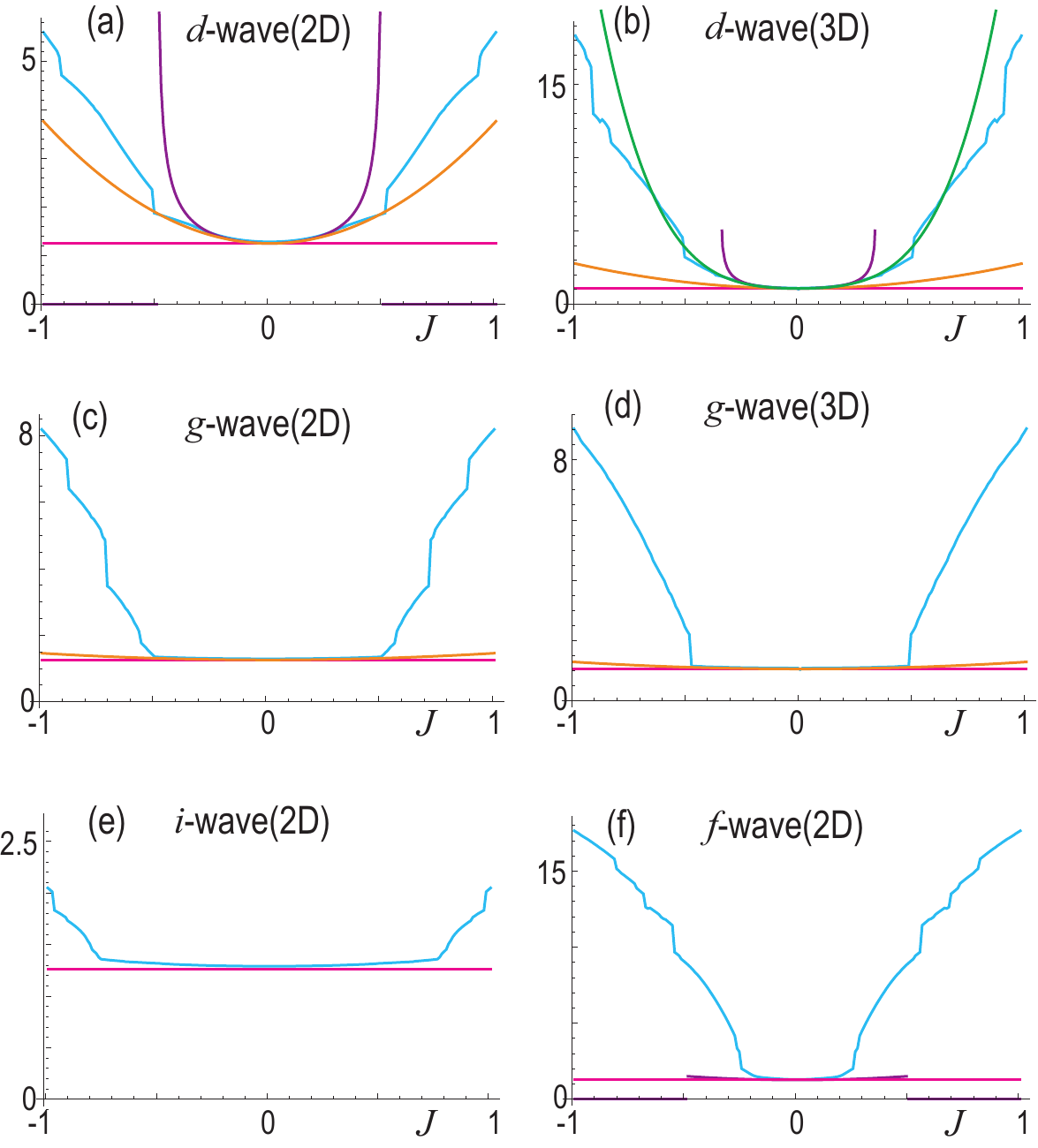}}
\caption{$J$ dependence of the Fermi volume $V^{\text{F}}$. (a) $d$-wave
magnet in 2D. (b) $d$-wave magnet in 3D. (c) $g$-wave magnet in 2D. (d) $g$%
-wave magnet in 3D. (e) $i$-wave magnet in 2D. (f) $f$-wave magnet in 2D.
Cyan curves represent numerical results based on tight-binding models.
Purple curves represent analytical results based on continuum models.
Magenta (orange, green) lines represent analytical results up to the zeroth
(second, fourth) order in $J$. The vertical axis is the Fermi volume in units of $2\pi m\mu /\hbar^2$ in two dimensions and $8\sqrt{2}\pi (m\mu)^{3/2}/(3\hbar^3)$ in three dimensions. The horizontal axis is $J$ in units of $\varepsilon _{0}$. We have set $\hbar ^{2}/\left( ma^{2}\right) =%
\protect\varepsilon _{0}/4$ and $\protect\mu =0.1\protect\varepsilon _{0}$.}
\label{FigArea}
\end{figure}

\begin{figure}[t]
\centerline{\includegraphics[width=0.48\textwidth]{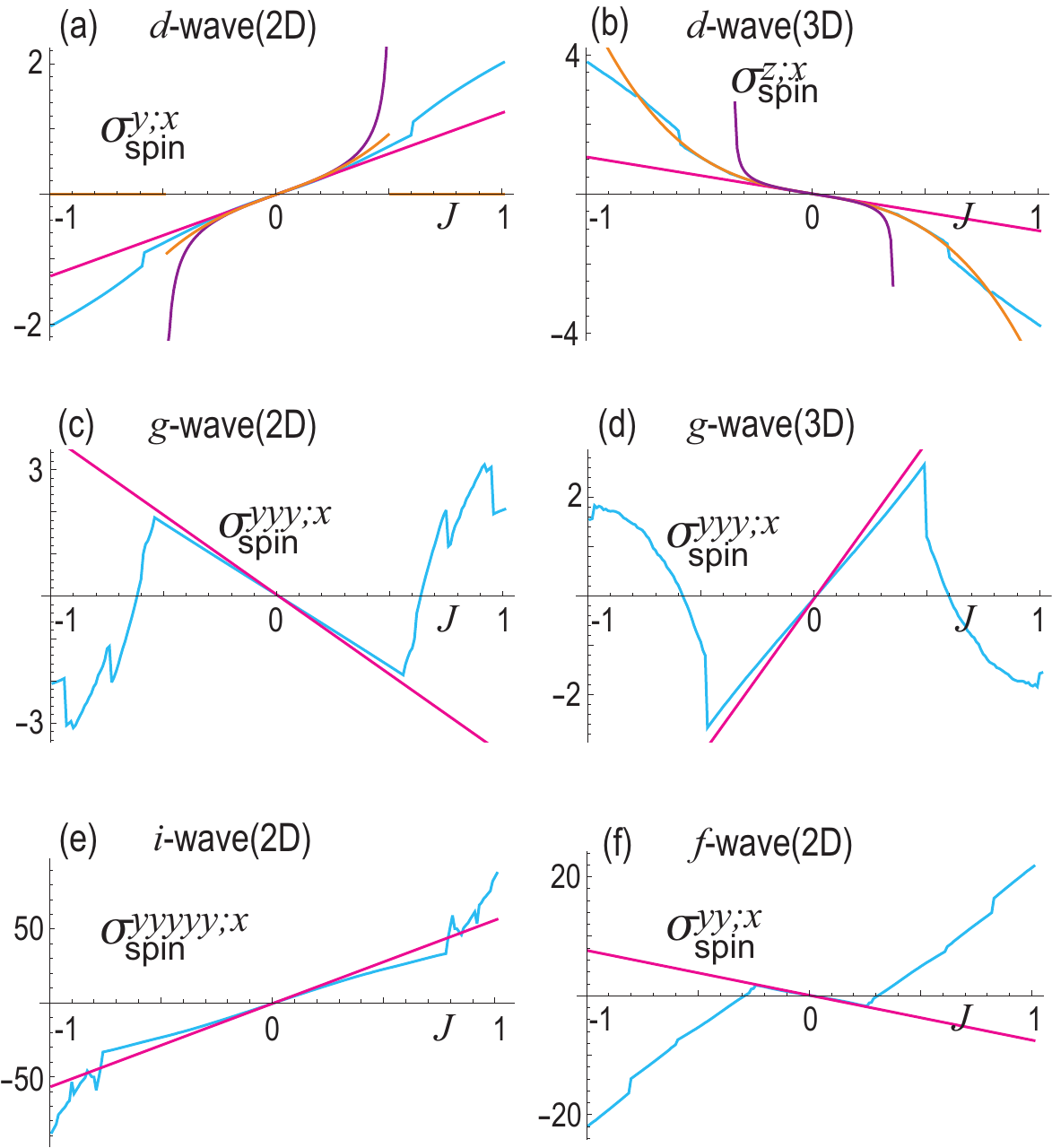}}
\caption{$J$ dependence of the spin-Drude conductivity. Cyan curves
represent numerical results based on tight-binding models. Purple curves
represent analytical results based on continuum models. Magenta (orange)
lines represent analytical results up to the zeroth (second) order in $J$. The horizontal axis is $J$ in units of $\varepsilon _{0}$. 
We have set $\hbar ^{2}/\left( ma^{2}\right) =\protect\varepsilon _{0}/4$
and $\protect\mu =0.1\protect\varepsilon _{0}$. See the caption of Fig.2
regarding others.}
\label{FigJCon}
\end{figure}

\section{$d$-wave altermagnet in 2D}

We analyze the Hamiltonian%
\begin{equation}
H^{\text{total}}=H_{0}^{2\text{D}}+H_{d}^{\text{2D}}.  \label{dModel}
\end{equation}%
The second term represents the $d$-wave altermagnet\cite%
{SmejRev,SmejX,SmejX2,Zu2023,Gho,Li2023,EzawaAlter,EzawaMetricC},%
\begin{equation}
H_{d}^{\text{2D}}=Jk_{x}k_{y}\sigma _{z}.
\end{equation}%
The spin-dependent energy is explicitly written as

\begin{equation}
\varepsilon _{s}=\frac{\hbar ^{2}\left( k_{x}^{2}+k_{y}^{2}\right) }{2m}%
+sJk_{x}k_{y}.  \label{D2DEne}
\end{equation}

We examine the condition (\ref{BasicCond}). We list all the nontrivial ones,%
\begin{align}
\frac{\partial \varepsilon _{s}}{\partial k_{x}} =&\frac{\hbar ^{2}k_{x}}{m}%
+sJk_{y},\quad \frac{\partial \varepsilon _{s}}{\partial k_{y}}=\frac{\hbar
^{2}k_{y}}{m}+sJk_{x},  \notag \\
\frac{\partial ^{2}\varepsilon _{s}}{\partial k_{x}^{2}} =&\frac{\partial
^{2}\varepsilon _{s}}{\partial k_{y}^{2}}=\frac{\hbar ^{2}}{m},\quad \frac{%
\partial ^{2}\varepsilon _{s}}{\partial k_{y}\partial k_{x}}=sJ,
\end{align}%
with $\varepsilon _{s}$ given by Eq.(\ref{D2DEne}), which could contribute
only to the spin conductivity for $\ell =0,1$. There is no $\ell $-th order
nonlinear spin-Drude conductivity for all $\ell \geq 2$. It is
straightforward to see that there is no spin conductivity for $\ell =0$.

We study the linear spin conductivity $\sigma _{\text{spin}}^{y;x}$ and $%
\sigma _{\text{spin}}^{x;x}$ corresponding to the choice of $\ell =1$. Note
that $\sigma _{\text{spin}}^{y;x}=\sigma _{\text{spin}}^{x;y}$ and $\sigma _{%
\text{spin}}^{x;x}=\sigma _{\text{spin}}^{y;y}$ due to the symmetry of the
Hamiltonian under the exchange of $k_{x}$ and $k_{y}$. We introduce the
polar coordinate $k_{x}=k\cos \phi $ and $k_{y}=k\sin \phi $.\ The Fermi
surface is analytically obtained as a function of $\phi $,%
\begin{equation}
\hbar k_{s}^{\text{F}}\left( \phi \right) =\sqrt{\frac{2\mu }{\frac{1}{m}%
+sJ\sin 2\phi }}.  \label{dphi}
\end{equation}%
It is shown for each spin $s$ in Fig.\ref{FigSurface}(c1). The Fermi volume
is give by%
\begin{equation}
V^{\text{F}}\equiv \int d^{2}{k}f_{s}^{(0)}=\frac{2\pi m\mu }{\hbar ^{2}%
\sqrt{1-J^{2}m^{2}}},  \label{D2SAna}
\end{equation}%
which is obtained with the use of (\ref{dphi}) for $\left\vert Jm\right\vert
<1$. This is the condition that the energy (\ref{D2DEne}) is positive for
sufficiently large momentum. The $J$\ dependence of the Fermi volume $V^{%
\text{F}}$ is shown in Fig.\ref{FigArea}(a).

First, we calculate the linear longitudinal spin-dependent conductivity $%
\sigma _{s}^{x;x}$. We obtain%
\begin{equation}
\sigma _{s}^{x;x}=\frac{e^{2}/\hbar ^{2}}{i\omega +1/\tau }\frac{\hbar ^{2}}{%
m}V^{\text{F}},
\end{equation}%
which yields $\sigma _{\text{spin}}^{x;x}=0$, because $\sigma _{s}^{x;x}$ is
independent of the spin $s$.

Next, we calculate the linear transverse spin-dependent conductivity $\sigma
_{s}^{y;x}$,%
\begin{equation}
\sigma _{s}^{y;x}=\sigma _{s}^{x;y}=s\frac{e^{2}/\hbar ^{2}}{i\omega +1/\tau 
}V^{\text{F}}J,
\end{equation}%
Hence, this spin conductivity is obtained as%
\begin{equation}
\sigma _{\text{spin}}^{y;x}=\sigma _{\text{spin}}^{x;y}=\frac{e^{2}/\hbar
^{2}}{i\omega +1/\tau }V^{\text{F}}J,  \label{sigD2}
\end{equation}%
with the use of the Fermi volume (\ref{D2SAna}). The $J$\ dependence of the $%
\sigma _{\text{spin}}^{y;x}$\ is shown in Fig.\ref{FigJCon}(a). We conclude
that there is only the linear spin conductivity.

The tight-binding model corresponding to the continuum model (\ref{dModel})
is given by 
\begin{equation}
H=\frac{1}{m}\left( 2-\cos k_{x}-\cos k_{y}\right) \sigma _{0}+J\sigma
_{z}\sin k_{x}\sin k_{y},
\end{equation}%
which is defined on the square lattice.

We calculate the spin-Drude conductivity $\sigma _{\text{spin}}^{y;x}$ based
on the continuum model and the tight-binding model, which are shown in Fig.%
\ref{FigEyyyx}(a). They agree each other very well in the vicinity of the
band bottom $\mu =0$. The analytical result of the Fermi volume (\ref{D2SAna}%
) well coincides with the tight-binding result for small $J$ as shown in Fig.%
\ref{FigArea}(a). The result (\ref{sigD2}) is only valid for $\left\vert
Jm\right\vert <1$ as shown in the purple curve in Fig.\ref{FigArea}(a).
However, by making a series expansion up to the second order in $J$, the
fitting becomes better as shown in the orange curve in Fig.\ref{FigArea}(a).
The $J$\ dependence of the $\sigma _{\text{spin}}^{y;x}$ is shown in Fig.\ref%
{FigJCon}(a), which is obtained numerically\ based on the tight-binding
model. The numerical and analytic results agree well for $\left\vert
Jm\right\vert <1$.

\section{$d$-wave altermagnet in 3D}

We consider the Hamiltonian of the $d$-wave altermagnet in three dimensions
given by%
\begin{equation}
H^{\text{total}}=H_{0}^{3\text{D}}+H_{d}^{3\text{D}}.  \label{d3Model}
\end{equation}%
The second term represents the $d$-wave altermagnet\cite{SmejX},%
\begin{equation}
H_{d}^{3\text{D}}=Jk_{z}\left( k_{x}+k_{y}\right) \sigma _{z}.
\end{equation}

The Fermi surface is analytically described in the spherical coordinate $%
k_{x}=k\cos \phi \sin \theta $, $k_{y}=k\sin \phi \sin \theta $ and $%
k_{z}=k\cos \theta $ as%
\begin{equation}
\hbar k_{s}^{\text{F}}\left( \phi ,\theta \right) =\sqrt{\frac{2\mu }{\frac{1%
}{m}+sJ\left( \cos \phi +\sin \phi \right) \sin 2\theta }},
\end{equation}%
which is shown in Fig.\ref{FigSurface}(c2). The Fermi volume is analytically
obtained as%
\begin{equation}
V^{\text{F}}=\frac{8\sqrt{2}\pi \left( m\mu \right) ^{3/2}}{3\hbar ^{3}\sqrt{%
1-2J^{2}m^{2}}}.  \label{FV3D}
\end{equation}

It follows from the condition (\ref{BasicCond}) that there is no $\ell $-th
order spin-Drude conductivity for all $\ell \geq 3$. It is straightforward
to see that there is no spin conductivity for $\ell =0$.

We calculate the linear longitudinal spin-dependent conductivity $\sigma
_{s}^{x;x}$. Because%
\begin{equation}
\frac{\partial ^{2}\varepsilon }{\partial k_{x}^{2}}=\frac{\partial
^{2}\varepsilon }{\partial k_{y}^{2}}=\frac{\partial ^{2}\varepsilon }{%
\partial k_{z}^{2}}=\frac{\hbar ^{2}}{m},
\end{equation}%
we obtain%
\begin{equation}
\sigma _{s}^{x;x}=\frac{e^{2}/m}{i\omega +1/\tau }V^{\text{F}}
\end{equation}%
with (\ref{FV3D}). This yields $\sigma _{\text{spin}}^{x;x}=0$, because $%
\sigma _{s}^{x;x}$ is independent of the spin $s$.

We next calculate the linear transverse spin-dependent conductivity $\sigma
_{s}^{z;x}$. Because%
\begin{equation}
\frac{\partial ^{2}\varepsilon _{s}}{\partial k_{z}\partial k_{x}}=\frac{%
\partial ^{2}\varepsilon _{s}}{\partial k_{y}\partial k_{z}}=sJ,
\end{equation}%
we obtain%
\begin{equation}
\sigma _{s}^{z;x}=\sigma _{s}^{x;z}=\sigma _{s}^{z;y}=\sigma _{s}^{y;z}=s%
\frac{e^{2}/\hbar ^{2}}{i\omega +1/\tau }V^{\text{F}}J.
\end{equation}%
Hence, the spin conductivity is obtained as%
\begin{equation}
\sigma _{\text{spin}}^{z;x}=\sigma _{\text{spin}}^{x;z}=\sigma _{\text{spin}%
}^{z;y}=\sigma _{\text{spin}}^{y;z}=\frac{e^{2}/\hbar ^{2}}{i\omega +1/\tau }%
V^{\text{F}}J,
\end{equation}%
The $J$\ dependence of the $\sigma _{\text{spin}}^{z;x}$\ is shown in Fig.%
\ref{FigJCon}(b). We conclude that there is only the linear spin
conductivity.

The tight-binding model corresponding to the continuum model (\ref{d3Model})
is given by 
\begin{align}
H=& \frac{\hbar ^{2}}{ma^{2}}\left( 3-\cos ak_{x}-\cos ak_{y}-\cos
ak_{z}\right) \sigma _{0}  \notag \\
& +J\sigma _{z}\sin ak_{z}\left( \sin ak_{x}+\sin ak_{y}\right) ,
\end{align}%
which is defined on the cubic lattice. We calculate the spin-Drude
conductivity based on the continuum model and the tight-binding model, which
are shown in Fig.\ref{FigEyyyx}(b). They agree each other very well in the
vicinity of the band bottom $\mu =0$. The Fermi surface is shown in Fig.\ref%
{FigSurface}(c2).

The $J$\ dependence of the Fermi volume $V_{s}$ is shown in Fig.\ref{FigArea}%
(b). The result (\ref{sigD2}) is only valid for $\left\vert Jm\right\vert <1/%
\sqrt{2}$ as shown in the purple curve in Fig.\ref{FigArea}(b). However, by
making a series expansion up to the fourth order in $J$, the fitting becomes
better as shown in the green curve in Fig.\ref{FigArea}(b).

The $J$\ dependence of the $\sigma _{\text{spin}}^{z;x}$\ is shown in Fig.%
\ref{FigJCon}(b). The series expansion up to the second order in $J$ well
fits the result based on the tight-binding model as shown in the orange
curve in Fig.\ref{FigJCon}(b).

\section{$f$-wave magnet in 2\textbf{D}}

A possibility of $f$-wave magnets was mentioned\cite{pwave,He}. The simplest
Hamiltonian of the $f$-wave magnet in two dimensions is given by%
\begin{equation}
H^{\text{total}}=H_{0}^{2\text{D}}+H_{f}^{2\text{D}}.  \label{fModel}
\end{equation}%
The second term represents the $f$-wave magnet,%
\begin{equation}
H_{f}=Jk_{x}\left( k_{x}^{2}-3k_{y}^{2}\right) \sigma _{z}.  \label{EqC}
\end{equation}%
The Fermi surface is described by%
\begin{equation}
\hbar k_{s}^{\text{F}}\left( \phi \right) =\sqrt{2\mu m}-2m^{2}\mu sJ\cos
3\phi ,
\end{equation}%
up to the first order in $J$. The Fermi surfaces $k_{s}^{\text{F}}\left(
\phi \right) $\ for up and down spins are shown in Fig.\ref{FigSurface}(d1).
Then, the Fermi volume is analytically obtained as%
\begin{equation}
V^{\text{F}}=\frac{2\pi m\mu }{\hbar ^{2}}\left( 1+\mu m^{3}J^{2}\right) ,
\label{VFF2D}
\end{equation}%
up to the second order in $J$. It is independent of the spin $s$.

We list all the nontrivial ones in the condition (\ref{BasicCond}),%
\begin{align}
\frac{\partial \varepsilon _{s}}{\partial k_{x}}=& \frac{\hbar ^{2}}{m}%
k_{x}+3sJ\left( k_{x}^{2}-k_{y}^{2}\right) ,  \notag \\
\frac{\partial \varepsilon _{s}}{\partial k_{y}}=& \frac{\hbar ^{2}}{m}%
k_{y}-6sJk_{x}k_{y}  \notag \\
\frac{\partial ^{2}\varepsilon _{s}}{\partial k_{x}^{2}}=& \frac{\hbar ^{2}}{%
m}+6sJk_{x},\qquad \frac{\partial ^{2}\varepsilon _{s}}{\partial k_{y}^{2}}=%
\frac{\hbar ^{2}}{m}-6sJk_{x},  \notag \\
\frac{\partial ^{2}\varepsilon _{s}}{\partial k_{x}\partial k_{y}}=&
-6sJk_{y},  \notag \\
\frac{\partial ^{3}\varepsilon _{s}}{\partial k_{x}\partial k_{y}^{2}}=&
-6sJ,\qquad \frac{\partial ^{3}\varepsilon _{s}}{\partial k_{x}^{3}}=6sJ.
\end{align}%
They could produce the spin conductivity (\ref{Drude}) only for $\ell =0,1,2$%
. There is no $\ell $-th order nonlinear spin-Drude conductivity for $\ell
\geq 3$.

We examine the spin conductivity for $\ell =0,1,2$. Calculating them
explicitly, we find that the nontrivial results are given only by the terms $%
\partial ^{3}\varepsilon _{s}/\partial k_{x}\partial k_{y}^{2}$ and $%
\partial ^{3}\varepsilon _{s}/\partial k_{x}^{3}$, yielding%
\begin{equation}
\int d^{2}kf_{s}^{(0)}\frac{\partial ^{3}\varepsilon _{s}}{\partial
k_{y}^{2}\partial k_{x}}=-\int d^{2}kf_{s}^{(0)}\frac{\partial
^{3}\varepsilon _{s}}{\partial k_{x}^{3}}=-6sV^{\text{F}}J,
\end{equation}%
where $V^{\text{F}}$ is the Fermi volume (\ref{VFF2D}). The second-order
nonlinear spin-Drude conductivity is given by%
\begin{align}
\sigma _{\text{spin}}^{yy;x}=& \sigma _{\text{spin}}^{xy;y}=6\frac{\left(
e/\hbar \right) ^{3}}{\left( i\omega +1/\tau \right) ^{2}}V^{\text{F}}J, 
\notag \\
\sigma _{\text{spin}}^{xx;x}=& -6\frac{\left( e/\hbar \right) ^{3}}{\left(
i\omega +1/\tau \right) ^{2}}V^{\text{F}}J.
\end{align}%
We conclude that there is only the second-order nonlinear spin-Drude
conductivity.

The tight-binding model corresponding to the continuum model (\ref{fModel})
is given by%
\begin{align}
H=& \frac{-2\hbar ^{2}}{3ma^{2}}\left( \sum_{j=1}^{3}\cos \left( a\mathbf{n}%
_{j}\cdot \mathbf{k}\right) -3\right) \sigma _{0}  \notag \\
& -4J\sigma _{z}\prod\limits_{j=1}^{3}\sin \left( a\mathbf{n}_{j}\cdot 
\mathbf{k}\right) .
\end{align}%
We calculate the spin-Drude conductivity based on the continuum model and
the tight-binding model, which are shown in Fig.\ref{FigEyyyx}(f). They
agree each other very well in the vicinity of the band bottom $\mu =0$.

The second-order nonlinear conductivity shows a nonreciprocity because the
current flowing direction is identical irrespective of the direction of
applied electric field. Especially, the nonreciprocity is perfect because
there is only the second-order nonlinear conductivity in the $f$-wave magnet.

The $J$\ dependence of the $\sigma ^{yy;x}$\ is shown in Fig.\ref{FigJCon}%
(f). The $\sigma ^{yy;x}$\ is linear in $J$\ in a wide range of $J$. The $J$%
\ dependence of the Fermi volume $V^{\text{F}}$ is shown in Fig.\ref{FigArea}%
(f). There is almost no $J$\ dependence in the Fermi volume for small $J$.

\section{$f$-wave magnet in 3D}

We consider the Hamiltonian of the $f$-wave magnet in three dimensions given
by%
\begin{equation}
H^{\text{total}}=H_{0}^{3\text{D}}+H_{f}^{3\text{D}}.  \label{f3D}
\end{equation}%
The second term represents the $f$-wave magnet,%
\begin{equation}
H_{f}^{3\text{D}}=Jk_{x}k_{y}k_{z}\sigma _{z}.
\end{equation}%
The Fermi surface is described by%
\begin{equation}
\hbar k_{s}^{\text{F}}\left( \phi ,\theta \right) =\sqrt{2\mu m}-\mu
m^{2}sJ\cos \theta \sin ^{2}\theta \sin 2\phi ,
\end{equation}%
up to the first order in $J$, which is a constant and independent of the
spin $s$. The Fermi volume is given by%
\begin{equation}
V^{\text{F}}=\frac{8\sqrt{2}\pi \left( m\mu \right) ^{3/2}}{3\hbar ^{3}}+%
\frac{16\sqrt{2}}{105\hbar ^{3}}\mu ^{2}J^{2}m^{4}\sqrt{m}\pi ,
\end{equation}%
up to the second order in $J$.

There is no $\ell $-th order spin-Drude conductivity for all $\ell \geq 3$.
It is straightforward to see that there is no spin conductivity for $\ell
=0,1$. We calculate the nonlinear spin-dependent conductivity with $\ell =2$%
.\ Because%
\begin{equation}
\frac{\partial ^{3}\varepsilon _{s}}{\partial k_{x}\partial k_{y}\partial
k_{z}}=sJ,
\end{equation}%
we obtain the nonlinear spin-Drude conductivity%
\begin{equation}
\sigma _{\text{spin}}^{zy;x}=-\frac{\left( e/\hbar \right) ^{3}}{\left(
i\omega +1/\tau \right) ^{2}}V^{\text{F}}J.
\end{equation}%
We conclude that there is only the second-order nonlinear spin-Drude
conductivity.

The tight-binding model corresponding to the continuum model (\ref{f3D}) is
given by 
\begin{align}
H=& \frac{\hbar ^{2}}{ma^{2}}\left( 3-\cos ak_{x}-\cos ak_{y}-\cos
ak_{z}\right) \sigma _{0}  \notag \\
& +J\sigma _{z}\sin ak_{x}\sin ak_{y}\sin ak_{z},
\end{align}%
which is defined on the cubic lattice. The Fermi surface is shown in Fig.\ref%
{FigSurface}(d2).

\section{$g$-wave altermagnet in 2D}

We consider the Hamiltonian of the $g$-wave altermagnet in two dimensions
given by%
\begin{equation}
H^{\text{total}}=H_{0}^{2\text{D}}+H_{g}^{2\text{D}}.  \label{gModel}
\end{equation}%
The second term represents the $g$-wave altermagnet\cite{SmejX},%
\begin{equation}
H_{g}^{2\text{D}}=Jk_{x}k_{y}\left( k_{x}^{2}-k_{y}^{2}\right) \sigma _{z}.
\label{EqD}
\end{equation}%
The spin-dependent energy leads%
\begin{equation}
\varepsilon _{s}=\frac{\hbar ^{2}\left( k_{x}^{2}+k_{y}^{2}\right) }{2m}%
+sJk_{x}k_{y}\left( k_{x}^{2}-k_{y}^{2}\right) .
\end{equation}%
The Fermi surface is described by%
\begin{equation}
\hbar k_{s}^{\text{F}}\left( \phi \right) =\sqrt{\frac{-1+\sqrt{1+4\mu
sJm^{2}\sin 4\phi }}{sJm\sin 4\phi }},
\end{equation}%
which is shown in Fig.\ref{FigSurface}(e1). The Fermi volume is obtained as%
\begin{equation}
V^{\text{F}}=\frac{2\pi m\mu }{\hbar ^{2}}\left( 1+\mu ^{2}m^{4}J^{2}\right)
,
\end{equation}%
up to the second order in $J$.

There is no $\ell $-th order nonlinear spin-Drude conductivity for all $\ell
\geq 4$. We then calculate explicitly the spin current for $\ell =0,1,2,3$.
The nontrivial conductivity arises\ only from the following terms,%
\begin{equation}
\frac{\partial ^{4}\varepsilon _{s}}{\partial k_{y}^{3}\partial k_{x}}%
=-6sJ,\qquad \frac{\partial ^{4}\varepsilon _{s}}{\partial k_{x}^{3}\partial
k_{y}}=6sJ.
\end{equation}%
They lead to the third-order nonlinear spin-Drude conductivity,%
\begin{align}
\sigma _{\text{spin}}^{yyy;x} =&\sigma _{\text{spin}}^{xyy;y}=-6\frac{\left(
e/\hbar \right) ^{4}}{\left( i\omega +1/\tau \right) ^{3}}V^{\text{F}}J, 
\notag \\
\sigma _{\text{spin}}^{xxx;y} =&\sigma _{\text{spin}}^{xxy;x}=6\frac{\left(
e/\hbar \right) ^{4}}{\left( i\omega +1/\tau \right) ^{3}}V^{\text{F}}J.
\label{g2Ds}
\end{align}%
We conclude that there is only the third-order nonlinear spin-Drude
conductivity.

The tight-binding model corresponding to the continuum model (\ref{gModel})
is given by 
\begin{align}
H_{0}^{2D}=& \frac{\hbar ^{2}}{ma^{2}}\left( 2-\cos ak_{x}-\cos
ak_{y}\right) \sigma _{0}  \notag \\
& +J\sin ak_{x}\sin ak_{y}\left( \cos ak_{x}-\cos ak_{y}\right) \sigma _{z},
\end{align}%
which is defined on the square lattice. We calculate the spin-Drude
conductivity based on the continuum model and the tight-binding model, which
are shown in Fig.\ref{FigEyyyx}(c). They agree each other very well in the
vicinity of the band bottom $\mu =0$.

The $J$\ dependence of the Fermi volume $V^{\text{F}}$ is shown in Fig.\ref%
{FigArea}(c). There is almost no $J$\ dependence in the Fermi volume for
small $J$. The $J$\ dependence of the $\sigma _{\text{spin}}^{yyy;x}$\ is
shown in Fig.\ref{FigJCon}(c). The$\ \sigma _{\text{spin}}^{yyy;x}$\ is
linear in $J$\ in a wide range of $J$. It means that the analytic result Eq.(%
\ref{g2Ds}) is applicable to a wide range of $J$.

\section{$g$-wave altermagnet in 3D}

We consider the Hamiltonian of the $g$-wave altermagnet in three dimensions
given by%
\begin{equation}
H^{\text{total}}=H_{0}^{3\text{D}}+H_{g}^{3\text{D}}.  \label{g3Model}
\end{equation}%
The second term represents the $g$-wave altermagnet in three dimensions\cite%
{SmejX},%
\begin{equation}
H_{g}^{\text{3D}}=Jk_{z}k_{x}\left( k_{x}^{2}-3k_{y}^{2}\right) \sigma _{z}.
\end{equation}%
The Fermi surface is described by%
\begin{equation}
\hbar k_{s}^{\text{F}}\left( \phi ,\theta \right) =\sqrt{\frac{-1+\sqrt{%
1+16\mu sJm^{2}\cos \theta \cos 3\phi \sin ^{3}\theta }}{sJm\sin ^{3}\theta
\cos \theta \cos 3\phi }},
\end{equation}%
which is shown $s$ in Fig.\ref{FigSurface}(e1). The Fermi volume is obtained
as%
\begin{equation}
V^{\text{F}}=\frac{8\sqrt{2}\pi \left( m\mu \right) ^{3/2}}{3\hbar ^{3}}+%
\frac{128\sqrt{2}\pi m^{11/2}\mu ^{7/2}J^{2}}{35\hbar ^{3}},
\end{equation}%
up to the second order in $J$. It is independent of the spin $s$.

There is no $\ell $-th order spin-Drude conductivity for all $\ell \geq 4$.
It is straightforward to see that there is no spin conductivity for $\ell
=0,1,2$.

We calculate the nonlinear spin-dependent conductivity with $\ell =3$.\
Because%
\begin{equation}
\frac{\partial ^{4}\varepsilon _{s}}{\partial k_{x}^{3}\partial k_{z}}=-%
\frac{\partial ^{4}\varepsilon _{s}}{\partial k_{x}\partial
k_{y}^{2}\partial k_{z}}=6sJ,
\end{equation}%
we obtain%
\begin{equation}
\sigma _{\text{spin}}^{xxx;z}=6\frac{\left( e/\hbar \right) ^{4}}{\left(
i\omega +1/\tau \right) ^{3}}V^{\text{F}}J.  \label{g3Ds}
\end{equation}%
We conclude that there is only the third-order nonlinear spin-Drude
conductivity.

The tight-binding model corresponding to the continuum model (\ref{g3Model})
is given by 
\begin{align}
H=& \left[ \frac{-2\hbar ^{2}}{3ma^{2}}\left( \sum_{j=1}^{3}\cos a\mathbf{n}%
_{j}\cdot \mathbf{k}-3\right) +\frac{\hbar ^{2}}{m_{z}a^{2}}\left( 1-\cos
ak_{z}\right) \right] \sigma _{0}  \notag \\
& -4J\sigma _{z}\sin ak_{z}\prod\limits_{j=1}^{3}\sin a\mathbf{n}_{j}\cdot 
\mathbf{k},
\end{align}%
where we have defined%
\begin{equation}
\mathbf{n}_{j}=\left( \cos \frac{2\pi j}{3},\sin \frac{2\pi j}{3}\right) ,
\end{equation}%
with $j=0,1,2$ and $\mathbf{k=}\left( k_{x},k_{y}\right) $. It is defined on
the layered triangular lattice. The Fermi surface is shown in Fig.\ref%
{FigSurface}(e2).

We calculate the spin-Drude conductivity based on the continuum model and
the tight-binding model, which are shown in Fig.\ref{FigEyyyx}(d). They
agree very well in the vicinity of the band bottom $\mu =0$.

The $J$\ dependence of the Fermi volume $V_{s}^{\text{F}}$ is shown in Fig.%
\ref{FigArea}(d). There is almost no $J$\ dependence in the Fermi volume for
small $J$. The $J$\ dependence of the $\sigma _{\text{spin}}^{xxx;z}$\ is
shown in Fig.\ref{FigJCon}(d). The $\sigma _{\text{spin}}^{xxx;z}$\ is
linear in $J$\ in a wide range of $J$. It means that the analytic result Eq.(%
\ref{g3Ds}) is applicable to a wide range of $J$.

\section{$i$-wave altermagnet in 2D}

We consider the Hamiltonian of the $i$-wave altermagnet in two dimensions
given by%
\begin{equation}
H^{\text{total}}=H_{0}^{2\text{D}}+H_{i}^{2\text{D}}.  \label{iModel}
\end{equation}%
The second term represents the $i$-wave altermagnet\cite{SmejX},%
\begin{equation}
H_{i}^{\text{2D}}=Jk_{x}k_{y}\left( 3k_{x}^{2}-k_{y}^{2}\right) \left(
k_{x}^{2}-3k_{y}^{2}\right) \sigma _{z}.
\end{equation}%
The Fermi surfaces for up and down spins are shown in Fig.\ref{FigSurface}%
(f1).

The Fermi surface is described by%
\begin{equation}
\hbar k_{s}^{\text{F}}\left( \phi \right) =\sqrt{2m\mu }\left( 1-2\mu
^{2}m^{3}sJ\sin 6\phi \right)
\end{equation}%
up to the first order in $J$. The Fermi volume is analytically obtained as%
\begin{equation}
V^{\text{F}}=\frac{2\pi m\mu }{\hbar ^{2}}\left( 1+\frac{\mu ^{4}m^{6}J^{2}}{%
2}\right) ,
\end{equation}%
up to the second order in $J$. It is independent of the spin $s$.

There is no $\ell $-th order spin-Drude conductivity for all $\ell \geq 6$.
It is straightforward to see that there is no spin conductivity for $\ell
=0,1,2,3,4$. We calculate the nonlinear spin-dependent conductivity for $%
\ell =5$.\ The nontrivial conductivity arises\ only from the following terms,%
\begin{equation}
\frac{\partial ^{6}\varepsilon _{s}}{\partial k_{y}^{5}\partial k_{x}}=-%
\frac{\partial ^{6}\varepsilon _{s}}{\partial k_{y}^{3}\partial k_{x}^{3}}=%
\frac{\partial ^{6}\varepsilon _{s}}{\partial k_{y}\partial k_{x}^{5}}=360sJ,
\end{equation}%
and those obtained by changing $x$ and $y$. The nontrivial spin-Drude
conductivity is given by%
\begin{equation}
\sigma _{\text{spin}}^{yyyyy;x}=\sigma _{\text{spin}}^{xxxxx;y}=-\sigma _{%
\text{spin}}^{xxxyy;y}=360\frac{\left( e/\hbar \right) ^{6}}{\left( i\omega
+1/\tau \right) ^{5}}V^{\text{F}}J.  \label{i2Ds}
\end{equation}%
We conclude that there is only the fifth-order nonlinear spin-Drude
conductivity.

The tight-binding model corresponding to the continuum model (\ref{iModel})
is given by 
\begin{align}
H=& \left[ \frac{-2\hbar ^{2}}{3ma^{2}}\left( \sum_{j=0}^{2}\cos \left( a%
\mathbf{n}_{j}^{\text{A}}\cdot \mathbf{k}\right) -3\right) \right.  \notag \\
& \left. +\frac{\hbar ^{2}}{m_{z}a^{2}}\left( 1-\cos ak_{z}\right) \right]
\sigma _{0}  \notag \\
& +16J\sigma _{z}\sin ak_{z}\prod\limits_{j=0}^{2}\sin \left( a\mathbf{n}%
_{j}^{\text{A}}\cdot \mathbf{k}\right) \prod\limits_{j=0}^{2}\sin \left( 
\sqrt{3}a\mathbf{n}_{j}^{\text{B}}\cdot \mathbf{k}\right) ,
\end{align}%
where we have defined%
\begin{align}
\mathbf{n}_{j}^{\text{A}}& =\left( \cos \frac{2\pi j}{3},\sin \frac{2\pi j}{3%
}\right) ,  \notag \\
\mathbf{n}_{j}^{\text{B}}& =\left( \sin \frac{2\pi j}{3},\cos \frac{2\pi j}{3%
}\right) ,
\end{align}%
with $j=0,1,2$. We calculate the spin-Drude conductivity based on the
continuum model and the tight-binding model, which are shown in Fig.\ref%
{FigEyyyx}(e). They agree each other very well in the vicinity of the band
bottom $\mu =0$.

The $J$\ dependence of the Fermi volume $V_{s}^{\text{F}}$ is shown in Fig.%
\ref{FigArea}(e). The $J$\ dependence of the $\sigma _{\text{spin}%
}^{yyyyy;z} $\ is shown in Fig.\ref{FigJCon}(e). The $\sigma _{\text{spin}%
}^{yyyyy;z}$\ is linear in $J$\ in a wide range of $J$. It means that the
analytic result Eq.(\ref{i2Ds}). is applicable to a wide range of $J$.

\section{$i$-wave altermagnet in 3D}

We consider the Hamiltonian of a $i$-wave altermagnet in three dimensions
given by%
\begin{equation}
H^{\text{total}}=H_{0}^{3\text{D}}+H_{i}^{\text{3D}}.
\end{equation}%
The first term represents the kinetic energy of free fermions\cite{SmejX},%
\begin{equation}
H_{i}^{\text{3D}}=J\left( k_{x}^{2}-k_{y}^{2}\right) \left(
k_{y}^{2}-k_{z}^{2}\right) \left( k_{z}^{2}-k_{x}^{2}\right) \sigma _{z}.
\end{equation}%
The Fermi surface is described by%
\begin{align}
\hbar k_{s}^{\text{F}}\left( \phi ,\theta \right) =& \sqrt{2m\mu }-\sqrt{%
2m\mu }8\mu ^{2}m^{3}sJ\sin ^{2}\theta \cos 2\phi  \notag \\
& \hspace{-20mm}\times \left( \cos ^{2}\theta -\sin ^{2}\theta \cos ^{2}\phi
\right) \left( \sin ^{2}\theta \sin ^{2}\phi -\cos ^{2}\theta \right) .
\end{align}%
up to the first order in $J$. The Fermi volume is analytically obtained as%
\begin{equation}
V^{\text{F}}=\frac{8\sqrt{2}\pi \left( m\mu \right) ^{3/2}}{3\hbar ^{3}}+%
\frac{16384\sqrt{2}\pi m^{15/2}\mu ^{11/2}J^{2}}{15015\hbar ^{3}},
\end{equation}%
up to the second order in $J$. It is independent of the spin $s$.

There is no $\ell $-th order spin-Drude conductivity for all $\ell \geq 6$.
It is straightforward to see that there is no spin conductivity for $\ell
=0,1,2,3,4$. We calculate the nonlinear spin-dependent conductivity for $%
\ell =5$.\ Because%
\begin{align}
&\frac{\partial ^{6}\varepsilon _{s}}{\partial k_{x}^{2}\partial k_{y}^{4}}=%
\frac{\partial ^{6}\varepsilon _{s}}{\partial k_{y}^{2}\partial k_{z}^{4}}=%
\frac{\partial ^{6}\varepsilon _{s}}{\partial k_{z}^{2}\partial k_{x}^{4}} 
\notag \\
&=-\frac{\partial ^{6}\varepsilon _{s}}{\partial k_{x}^{4}\partial k_{y}^{2}}%
=-\frac{\partial ^{6}\varepsilon _{s}}{\partial k_{y}^{4}\partial k_{z}^{2}}%
=-\frac{\partial ^{6}\varepsilon _{s}}{\partial k_{z}^{4}\partial k_{x}^{2}}%
=48sJ,  \label{5thDrude}
\end{align}%
\emph{\ }we obtain%
\begin{equation}
\sigma _{\text{spin}}^{yyyx;x}=48\frac{\left( e/\hbar \right) ^{6}}{\left(
i\omega +1/\tau \right) ^{5}}V^{\text{F}}J,
\end{equation}%
and similar ones arising from Eq.(\ref{5thDrude}). We conclude that there is
only the fifth-order nonlinear spin-Drude conductivity.

The tight-binding model corresponding to the continuum model (\ref{f3D}) is
given by 
\begin{align}
H=& \frac{\hbar ^{2}}{ma^{2}}\left( 3-\cos ak_{x}-\cos ak_{y}-\cos
ak_{z}\right) \sigma _{0}  \notag \\
& +8J\sigma _{z}\left( \cos ak_{y}-\cos ak_{x}\right) \left( \cos
ak_{z}-\cos ak_{y}\right)  \notag \\
& \times \left( \cos ak_{x}-\cos ak_{z}\right) ,
\end{align}%
which is defined on the cubic lattice. The Fermi surface is shown in Fig.\ref%
{FigSurface}(f2).

\section{Discussion}

We studied the higher-order nonlinear spin-Drude conductivity in higher-wave
symmetric magnets characterized by the number of nodes. We found that the
system having ($\ell +1$) nodes has only the $\ell $-th order nonlinear
spin-Drude conductivity. It is highly contrasted to the usual system\cite%
{Ideue}, where various higher-order contributions are present. Especially,
only the third-order nonlinear conductivity is present in $g$-wave
altermagnets and only the fifth-order nonlinear conductivity is present in $%
i $-wave altermgnets. The prominent feature is that $f$-wave magnets in two
dimensions generate the perfect nonreciprocal spin current because there is
only the second-order nonlinear spin current.

These results are derived from the two-band models, where the pseudospin
degrees of freedom are absent, and only a certain order of the nonlinear
spin current is generated. However, there are systems involving the
pseudospin degrees of freedom, where the Hamiltonian is not diagonal in
general. The study of such systems is a future problem.

The $d$-wave magnet in two dimensions is realized in organic materials\cite%
{Naka}, perovskite materials\cite{NakaB}, and twisted magnetic Van der Waals
bilayers\cite{YLiu}. The $d$-wave altermagnet in three dimensions is
realized in RuO$_{2}$\cite{Ahn,SmeRuO,Tsch,Fed,Lin}, Mn$_{5}$Si$_{3}$\cite%
{Leiv} and FeSb$_{2}$\cite{Mazin}. A $g$-wave altermagnet is realized in
twisted magnetic Van der Waals bilayers\cite{YLiu}. The Fermi surface
splitting of the $g$-wave altermagnet in three dimensions is experimentally
observed in MnTe\cite{Krem,LeeG,Osumi,Haj,Lee,Masuda} and CrSb\cite%
{Reim,GYang,Ding,Cli,WLu}. A $i$-wave altermagnet in two dimensions is
realized in twisted magnetic Van der Waals bilayers\cite{YLiu}. It was
theoretically proposed that CeNiAsO is a $p$-wave magnet\cite{pwave}. Our
work will motivate a further search on materialization of higher-wave
symmetric magnets.

We estimate the magnitude of electric field. The dimensionless electric
field is given by $E/\left( \frac{\hbar k}{e\tau }\right) $ with 
\begin{equation}
\frac{\hbar k}{e\tau }=6.6\times 10^{5}\left[ \text{kgm/As}^{3}\right] ,
\end{equation}%
where we have used the relaxation time\cite{Trama,Du} $\tau =3.4\times
10^{-12}$s and the typical momentum\cite{Krem} $k=0.35$\AA $^{-1}$.
Experimental value of electric field\cite{Sala} is $E=1$[V/$\mu $m]$=10^{6}$%
[kgm/As$^{3}$]. It is possible to apply larger electric field\cite{PHe} $%
E=10 $[V/$\mu $m] satisfying $E/\left( \frac{\hbar k}{e\tau }\right) >1$,
where the higher-order nonlinear conductivity is significantly large.
High-efficiency spin-current generation proportional to $E^{\ell }$\ is
possible in them if we apply electric field larger than $E=6.6\times 10^{5}%
\left[ \text{kgm/As}^{3}\right] $.

The author is very much grateful to M. Hirschberger for helpful discussions
on the subject. This work is supported by CREST, JST (Grants No. JPMJCR20T2)
and Grants-in-Aid for Scientific Research from MEXT KAKENHI (Grant No.
23H00171).


\end{document}